\documentclass{webofc}

\usepackage[varg]{txfonts}   
\usepackage{hyperref}
\usepackage{url}

\usepackage{amsmath}

\hypersetup{colorlinks=true,citecolor=blue,urlcolor=blue,linkcolor=blue}
\usepackage{titlesec}

\titlespacing*{\section}{0pt}{1ex plus 0.5ex minus 0.2ex}{0.8ex plus 0.3ex}
\titlespacing*{\subsection}{0pt}{0.8ex plus 0.3ex minus 0.2ex}{0.5ex plus 0.2ex}

\usepackage{lineno}

\makeatletter
\g@addto@macro\normalsize{%
  \setlength\abovedisplayskip{4pt}%
  \setlength\belowdisplayskip{4pt}%
  \setlength\abovedisplayshortskip{4pt}%
  \setlength\belowdisplayshortskip{4pt}%
}
\makeatother

\begin{document}%
%
%

%
\title{Probing jet hadrochemistry modification with measurements of $\mathbf{\pi}$, K, and p in jets and the underlying event in pp and \mbox{Pb--Pb} collisions at $\mathbf{\sqrt{\textit{s}_{\textup{NN}}} = 5.02}$ TeV}
%
%

\author{\firstname{Sierra} \lastname{Cantway}\inst{1}\thanks{\email{sierra.lisa.weyhmiller@cern.ch}} for the ALICE Collaboration
}

\institute{Yale University Wright Laboratory 272 Whitney Ave., New Haven, USA}

\abstract{Measurements of jet substructure observables in heavy-ion (HI) collisions provide powerful constraints on the microscopic mechanisms of interactions between energetic partons and the quark--gluon plasma (QGP). Although there has been remarkable progress in measuring inclusive jet substructure, a complete understanding of identified particle production inside jets (jet hadrochemistry) and its modification in HI collisions remains elusive. Jet quenching models predict that the jet hadrochemical composition is modified in the QGP, arising from both jet-medium interactions and altered particle production in the jet wake. Measurements of jet hadrochemistry can help discriminate between proposed jet-medium interaction mechanisms. Enabled by the excellent particle identification (PID) capabilities of ALICE, we present the first measurements of $\pi$, K, and p ratios within jets and the underlying event as a function of particle transverse momentum in pp and \mbox{Pb--Pb} collisions at $\sqrt{s_{\rm{NN}}}$ = 5.02 TeV. These measurements provide insight into soft particle production mechanisms and helps distinguish various jet quenching effects.}
\maketitle
\section{Introduction}
\label{intro}
Jets are modified from a vacuum baseline through interactions with a QGP medium. Jets are also expected to modify the medium itself by inducing a correlated medium response, resulting in an enhancement of particles in the jet-going direction, particularly at large angles~\cite{WakeKrishna}. Measurements of the internal structure of these jets (jet substructure) can constrain these QGP-parton interactions. Many jet substructure measurements focus on inclusive charged hadrons in jets~\cite{JetReview2018}, leaving knowledge of the identified particle composition of jets  (jet hadrochemistry) and its modification in the QGP incomplete. However, AMPT simulations predict that jet hadrochemistry can be modified in heavy-ion (HI) collisions due to parton coalescence in the jet wake~\cite{Luo:2023}. Conversely, theoretical calculations predict modified particle composition due to enhanced parton splittings in the medium~\cite{Sapeta:2007ad}. Therefore, jet hadrochemistry measurements could be sensitive to hadronization mechanisms, modified jet fragmentation, and wake contributions. In these proceedings, the ALICE collaboration presents measurements of the K/$\pi$ and p/$\pi$ ratios in jets and the underlying event (UE) in pp and \mbox{Pb--Pb} collisions to probe these effects.

\section{Experimental Details}
\label{experiment}
This analysis is performed on 0$-$10\% \mbox{Pb--Pb} and minimum-bias pp collision data at $\sqrt{s_{\rm{NN}}} = 5.02$ TeV. In \mbox{Pb--Pb}, the jet $\textit{p}_{\mathrm{T}}$ is corrected for the average event-wise background with the pedestal subtraction method~\cite{RhoSub}, requiring a jet $\textit{p}_{\mathrm{T}}$ after pedestal subtraction ($p^{\rm{raw~sub}}_{\rm{T,~ch~jet}}$) of 60\textendash140 GeV/$c$. 

To remove remaining background constituents inside the jets, PID is performed on different particle sources, i.e. the measurement regions correlated to how a particle is produced. These are then used to subtract the UE using the perpendicular cone method~\cite{PerpConeOG} in both pp and \mbox{Pb--Pb} collision data. The jet cone particle source (JC) is defined as all particles inside the reconstructed jet cones, i.e., both the jet signal and UE particles in the cone. The perpendicular cone particle source (PC) consists of all particles inside $R=0.4$ cones at $\Delta \varphi =90^{\circ}$ and $\Delta \eta =0$ from the jet axis. The inclusive particle production (without any jet requirements) is also measured for comparison.

PID is performed via fits to the Time-Of-Flight (TOF) $n_\sigma$ particle hypothesis distributions. The raw particle yield as a function of $\textit{p}_{\mathrm{T}}$ is obtained by integrating these fits for each species. Standard $\textit{p}_{\mathrm{T}}$-dependent, bin-by-bin PID corrections (tracking efficiency, TOF matching efficiency, and primary fraction) are then performed~\cite{ALICEIncPID}. 

Before performing the subtraction, the particle yields are normalized by area, i.e.,
\begin{equation}
\frac{\rm{d}\rho}{\rm{d}\textit{p}_{\mathrm{T}}^{\rm{track}}} = \frac{1}{N_{\rm{trig~evt}}} \frac{1}{A_{\rm{acc}}} \frac{\rm{d}\it{N}}{\rm{d}\textit{p}_{\mathrm{T}}^{\rm{track}}},
\label{eq:density}
\end{equation}
where $N_{\rm{trig~evt}}~A_{\rm{acc}}$ is the total acceptance area for triggered events for each particle source, and $\frac{\rm{d}\it{N}}{\rm{d}\textit{p}_{\mathrm{T}}^{\rm{track}}}$ is the particle yield. The PC yield is used to compare the jet and UE yields and to subtract the UE from the JC. 
Jets are more likely to be reconstructed on an upward fluctuation of the UE such that the PC alone underestimates the JC background. A scaling factor ($c_{\rm{UE~bias}}$) is applied to the PC yields to correct for this effect. In pp data, the UE is small, so $c_{\rm{UE~bias}}=1$. For a more detailed description of the analysis method, see Ref.~\cite{SierraHP24Proceed}. 
 



The jet $\textit{p}_{\mathrm{T}}$ smearing from detector and background effects is much smaller in pp than \mbox{Pb--Pb} due to the large UE in HI collisions. To enable comparisons between \mbox{Pb--Pb} and pp jets at the reconstructed level, the pp spectra must have their jet $\textit{p}_{\mathrm{T}}$ smeared to \mbox{Pb--Pb} conditions, which may also affect the particle $\textit{p}_{\mathrm{T}}$ spectra. To do this, the rate at which each pp $p^{\rm{raw}}_{\rm{T,~ch~jet}}$ interval contributes to the measured $p^{\rm{raw~sub}}_{\rm{T,~ch~jet}}$ interval in \mbox{Pb--Pb} conditions is determined. First, random cones placed in inclusive \mbox{Pb--Pb} data events are used to obtain the distribution of residual pure UE fluctuations after pedestal subtraction ($\delta \textit{p}_{\mathrm{T}} = \Sigma p_{\rm{T}}^{track} \rm{~in ~cone} - \rho*$$A_{\rm{cone}}$). The effects of these pure UE fluctuations are then propagated to the jets by randomly sampling the $\delta \textit{p}_{\mathrm{T}}$ distribution and the raw measured pp jet spectra to create a hybrid jet whose $p^{\rm{raw~sub}}_{\rm{T,~ch~jet}}$ is the sampled $\delta \textit{p}_{\mathrm{T}}$ value plus the sampled pp $p^{\rm{raw}}_{\rm{T,~ch~jet}}$ value. If the hybrid $p^{\rm{raw~sub}}_{\rm{T,~ch~jet}}$ is between 60-140 GeV/$c$, its associated pp data jet $\textit{p}_{\mathrm{T}}$ enters a response matrix. Normalizing this response by the total number of hybrid jets gives the rate at which each pp data jet $\textit{p}_{\mathrm{T}}$ interval contributes to these "\mbox{Pb--Pb} like" hybrid jets. The data pp particle spectra for each $p^{\rm{raw}}_{\rm{T,~ch~jet}}$ interval are then combined according to these fractions to produce smeared pp distributions, which represent the expected \mbox{Pb--Pb} measurement in the absence of quenching. The effect of different detector conditions between the pp and \mbox{Pb--Pb} datasets on the smeared pp jet $\textit{p}_{\mathrm{T}}$ is included as a systematic uncertainty. The smeared pp spectra can be directly compared to the measured \mbox{Pb--Pb} distributions after UE subtraction to probe the effects of jet quenching. 

It is possible that jet background smearing could affect each track $\textit{p}_{\mathrm{T}}$ interval differently. A systematic uncertainty for this effect is included in the folded result by widening the $\delta \textit{p}_{\mathrm{T}}$ fluctuations by the variation of the jet energy resolution with track $\textit{p}_{\mathrm{T}}$. The response matrix is recalculated for each track $\textit{p}_{\mathrm{T}}$ with this wider $\delta \textit{p}_{\mathrm{T}}$ fluctuation distribution. For each track $\textit{p}_{\mathrm{T}}$ interval, the measured data pp particle yield is then combined for each of these pp jet $\textit{p}_{\mathrm{T}}$ intervals according to these fractions, giving an alternative smeared pp jet track spectra.

\section{Results}
\label{results}



\begin{figure}[!h]
  \centering
  \includegraphics[width=0.495\textwidth,clip]{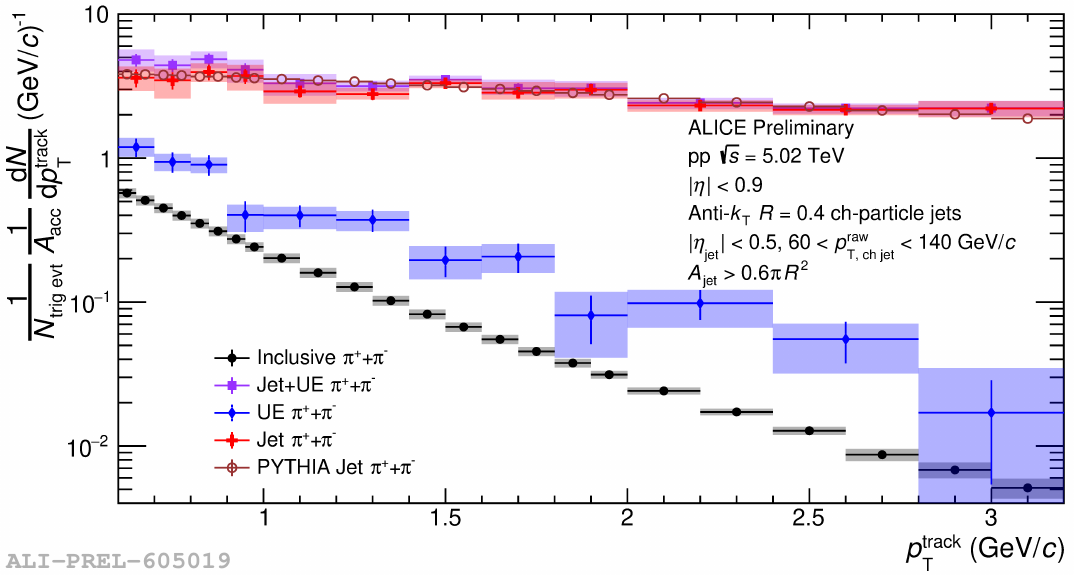} \\
  \includegraphics[width=0.495\textwidth,clip]{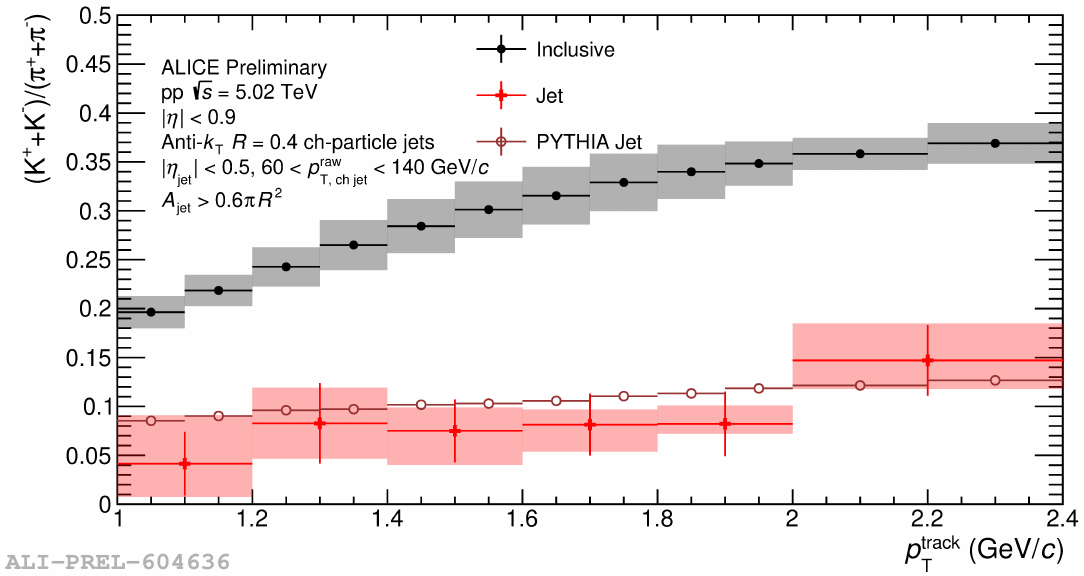}
  \includegraphics[width=0.495\textwidth,clip]{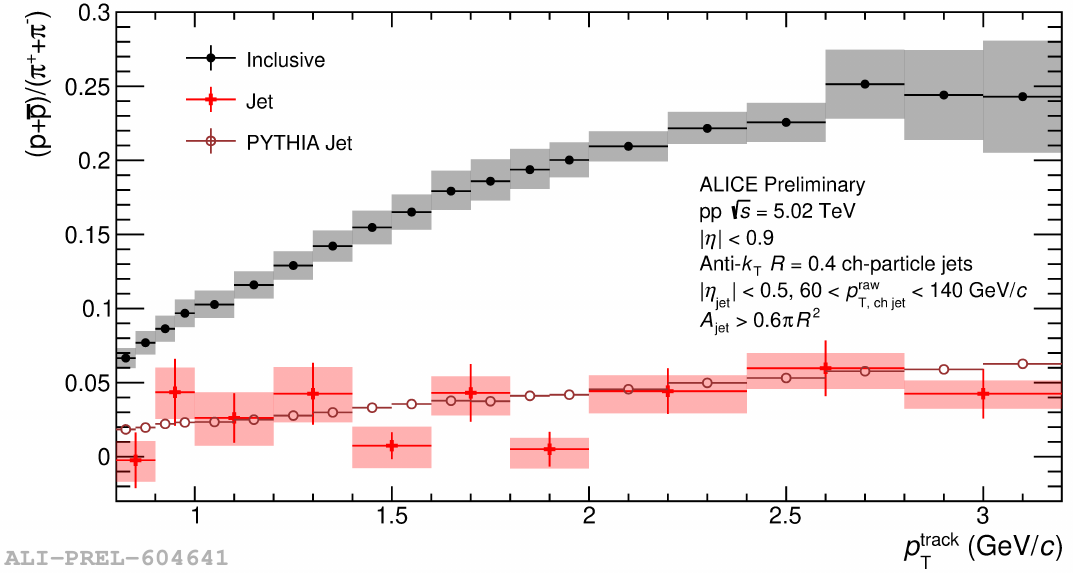}

\caption{Top: The $\pi$ spectra for inclusive, jet+UE, UE, and jet signal in pp collisions. Bottom: The K/$\pi$ (left) and p/$\pi$ (right) particle ratios in jets and inclusive particles in pp collisions.}
\label{fig:pp_plots}
\end{figure}

 The inclusive and jet+UE spectra are given directly by $\frac{\rm{d}\rho_{\rm{inc}}}{\rm{d}\textit{p}_{\mathrm{T}}^{\rm{track}}}$ and $\frac{\rm{d}\rho_{\rm{JC}}}{\rm{d}\textit{p}_{\mathrm{T}}^{\rm{track}}}$, respectively. The UE and jet spectra are given by 
\begin{align}
\frac{\rm{d}\rho_{\rm{UE}}}{\rm{d}\textit{p}_{\mathrm{T}}^{\rm{track}}} &= \frac{\rm{d}\rho_{\rm{PC}}}{\rm{d}\textit{p}_{\mathrm{T}}^{\rm{track}}} c_{\rm{UE~bias}}, \\
\frac{\rm{d}\rho_{\rm{Jet}}}{\rm{d}\textit{p}_{\mathrm{T}}^{\rm{track}}} &= \frac{\rm{d}\rho_{\rm{Jet+UE}}}{\rm{d}\textit{p}_{\mathrm{T}}^{\rm{track}}} - \frac{\rm{d}\rho_{\rm{UE}}}{\rm{d}\textit{p}_{\mathrm{T}}^{\rm{track}}}.
\end{align} 
Figure~\ref{fig:pp_plots} shows the $\pi$ spectra and species ratios in pp collisions. The corresponding distributions in \mbox{Pb--Pb} collisions are discussed in Ref.~\cite{SierraHP24Proceed}. The UE yields are observed to be denser than the inclusive yields due to the jet pedestal effect~\cite{JetPedestal}. The Jet+UE is dominated by the jet portion, and the jet portion of the Jet+UE yields becomes fractionally larger as $\textit{p}_{\mathrm{T}}$ increases. Similar trends are observed for the K and p spectra. The K/$\pi$ and p/$\pi$ ratios are lower in jets than in the inclusive case, indicating that there is less baryon and strangeness production in jets relative to inclusive production. PYTHIA describes the pp jet K/$\pi$ and p/$\pi$ ratios well. 

\begin{figure}[!h]
\centering
\includegraphics[width=0.495\textwidth,clip]{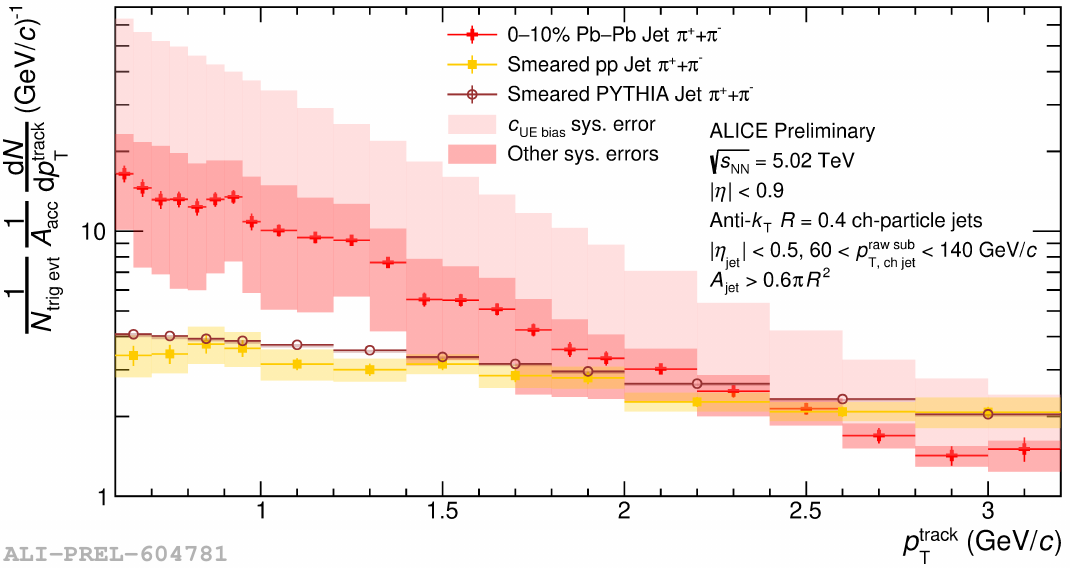}
\includegraphics[width=0.495\textwidth,clip]{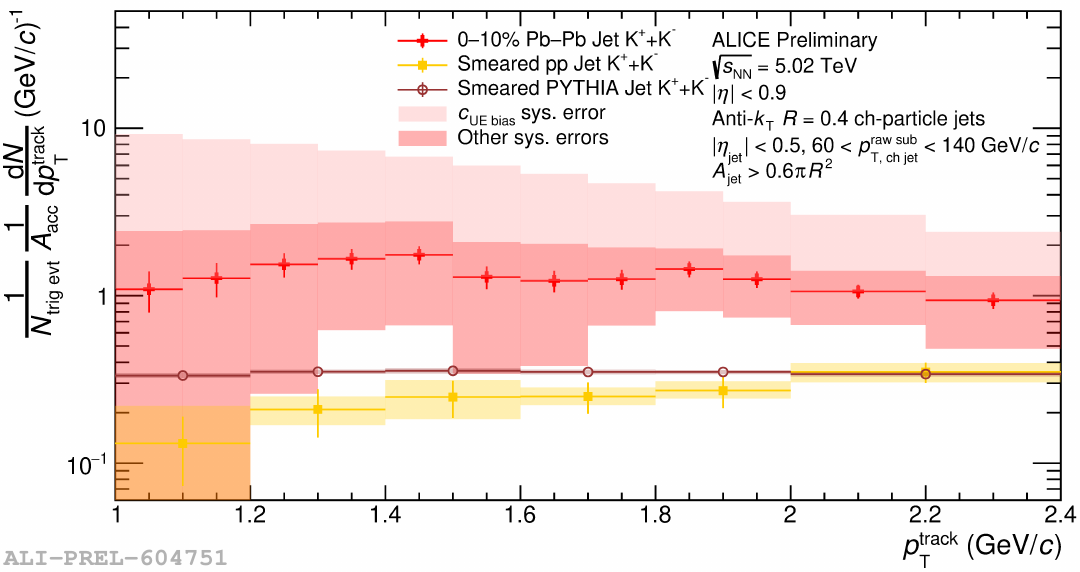}
\includegraphics[width=0.495\textwidth,clip]{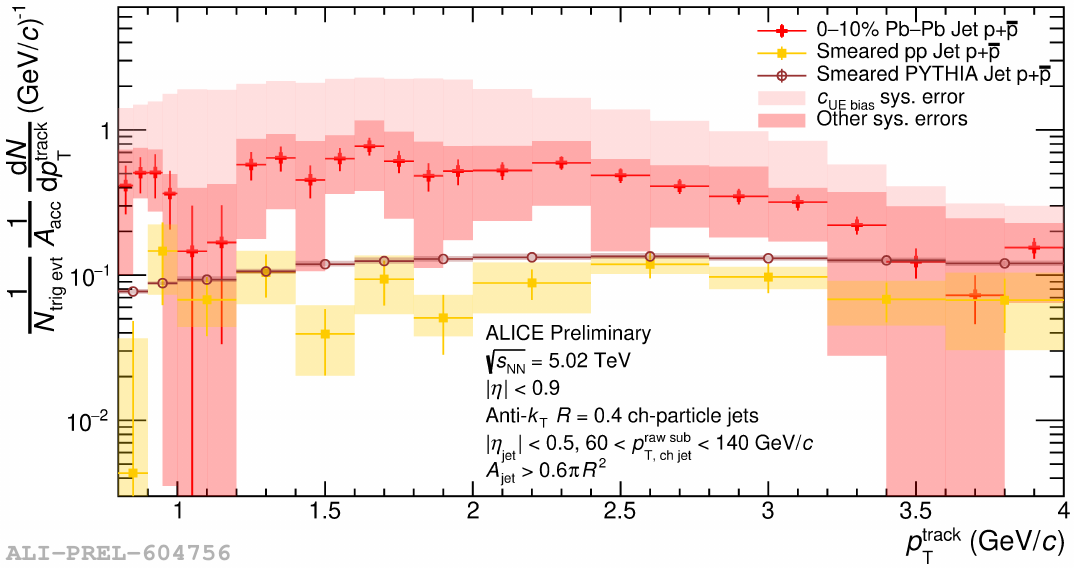}

\caption{The $\pi$ (top left), K (top right), and p (bottom) spectra in 0$-$10\% \mbox{Pb--Pb} and smeared pp jets.}
\label{fig:pi yields smear}
\end{figure} 



\begin{figure}[!h]
  \centering
  \includegraphics[width=0.495\textwidth,clip]{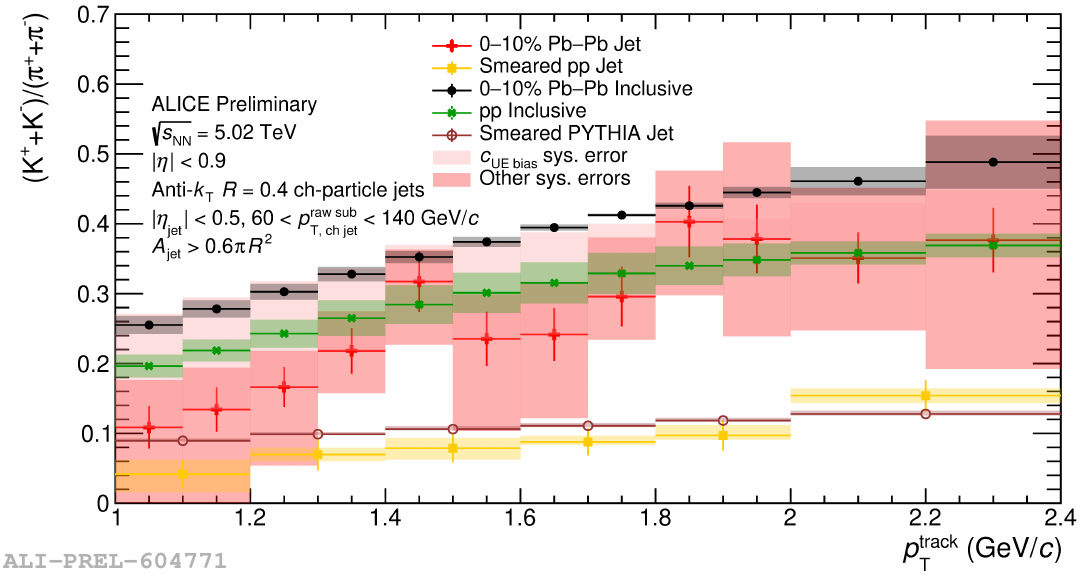}
  \includegraphics[width=0.495\textwidth,clip]{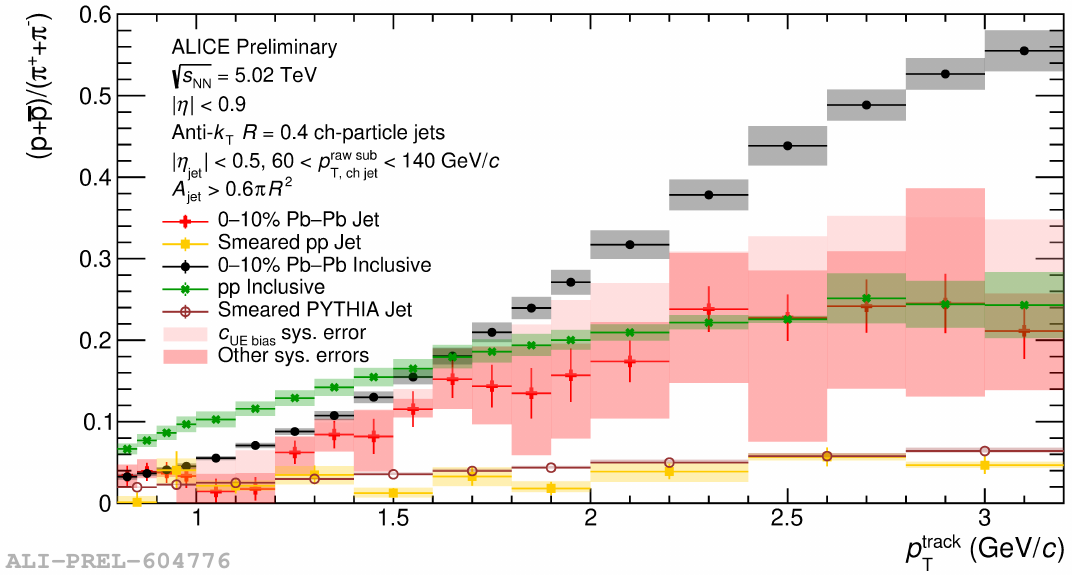}

\caption{The K/$\pi$ (left) and p/$\pi$ (right) particle ratios in 0$-$10\% \mbox{Pb--Pb} and smeared pp jets.}
\label{piKp_ratios_smearpp_PbPb}
\end{figure}

  The comparisons between the \mbox{Pb--Pb} and smeared pp jet $\pi$, K, and p spectra are shown in Fig.~\ref{fig:pi yields smear}. There is a hint of enhancement of the low $\textit{p}_{\mathrm{T}}$ $\pi$ yield and a hint of suppression of intermediate $\textit{p}_{\mathrm{T}}$ $\pi$ yield in \mbox{Pb--Pb} jets compared to pp jets. These conclusions corroborate previous species-inclusive jet fragmentation modification findings~\cite{CMSfrag_fxn, ATLAS_frag_fxn}. Hints of enhancement of intermediate $\textit{p}_{\mathrm{T}}$ K and p in \mbox{Pb--Pb} jets compared to pp jets are also observed. The comparison between the smeared pp and \mbox{Pb--Pb} jet species ratios is shown in Fig.~\ref{piKp_ratios_smearpp_PbPb}. Hints of strangeness and baryon enhancement are seen in \mbox{Pb--Pb} jets compared to pp jets. These results are the first hints of jet hadrochemistry modification in HI collisions, qualitatively agreeing with the enhancement of jet K/$\pi$ and p/$\pi$ in medium compared to vacuum predicted by theory~\cite{Luo:2023, Sapeta:2007ad}. 

\section{Conclusions}
\label{conclusions}

These proceedings present measurements of $\pi$, K, p production in jets and UE in pp and \mbox{Pb--Pb} collisions. These results hint at increasing baryon and strangeness production from pp jets to \mbox{Pb--Pb} jets, and then again from \mbox{Pb--Pb} jets to \mbox{Pb--Pb} UE. These results comprise the first hints of jet hadrochemistry modification in HI collisions and are in qualitative agreement with theory predictions. The next step is to unfold both the pp and \mbox{Pb--Pb} jet measurements to enable quantitative comparisons with theory, which could differentiate between jet hadrochemistry modification due to medium response and modified jet fragmentation. Also, the PID particle $\textit{p}_{\mathrm{T}}$ range will be extended and the centrality dependence explored. Finally, a differential analysis of the radial dependence of the jet particle ratios is planned, as the wake response is expected to increase the jet hadrochemistry modification at larger angles~\cite{Luo:2023}. 

%
\bibliography{refs}

\end{document}